\documentclass[twocolumn,prb,showpacs,aps, superscriptaddress]{revtex4}
\usepackage{graphicx}
\usepackage{ulem}

\begin{document}
\bibliographystyle{prsty}
\title{Dynamics of electronic transport in a semiconductor superlattice with a shunting side layer}

\author{ Huidong Xu}
\affiliation{Duke University, Department of Physics, Box 90305, Durham, NC 27708-0305, USA}

\author{Andreas Amann}
\affiliation{Tyndall National Institute, Lee Maltings, Cork, Ireland}

\author{Eckehard Sch{\"{o}}ll}
\affiliation{Institut f{\"{u}}r Theoretische Physik, Technische Universit{\"{a}}t Berlin, Hardenbergstra{\ss}e 36, 10623, Berlin, Germany }

\author{Stephen W. Teitsworth}
\affiliation{Duke University, Department of Physics, Box 90305, Durham, NC 27708-0305, USA}

\date{ \today }
\begin{abstract}
We study a model describing electronic transport in a weakly-coupled
semiconductor superlattice with a shunting side layer. Key
parameters include the lateral size of the superlattice, the
connectivity between the quantum wells of the superlattice and the
shunt layer, and the conduction properties of the shunt layer.  For a superlattice with small lateral extent and high
quality shunt, static electric field domains are \textit{suppressed} and a
spatially-uniform field configuration is predicted to be stable, a
result that may be useful for proposed devices such as a
superlattice-based TeraHertz (THz) oscillators.  As the lateral size
of the superlattice increases, the uniform field configuration loses
its stability to either static or dynamic field domains, regardless
of shunt properties. A lower quality shunt generally leads to
regular and chaotic current oscillations and complex spatio-temporal
dynamics in the field profile. Bifurcations separating static and
dynamic behaviors are characterized and found to be dependent on the
shunt properties.

\end{abstract}

\maketitle

\section{Introduction}
Theoretical work by Esaki and Tsu \cite{EsakiT70} in 1970 was the
first to propose a Bloch oscillator based on a superlattice (SL)
structure. In that paper, they derived current-voltage (I-V)
characteristics of a SL which showed negative differential
conductivity (NDC) associated with Bloch oscillations\cite{Bloch28,Zener34} of the miniband electrons under a DC bias.
However, direct observation of Bloch oscillations is difficult due
to decoherence caused by electron scattering. In other important
early work, Ktitorov, Simin and Sindalovskii \cite{KtitorovS72} predicted a negative
high-frequency differential conductivity and associated
amplification of high frequency signals thereby suggesting an
alternative means of THz oscillation. This dynamic conductivity
remains negative up to the Bloch frequency $\omega_B$ and reaches a
resonance minimum at a frequency closely below $\omega_B$,
suggesting that the SL may serve as an active medium for THz
radiation.

\begin{figure}
\includegraphics[width=7cm]{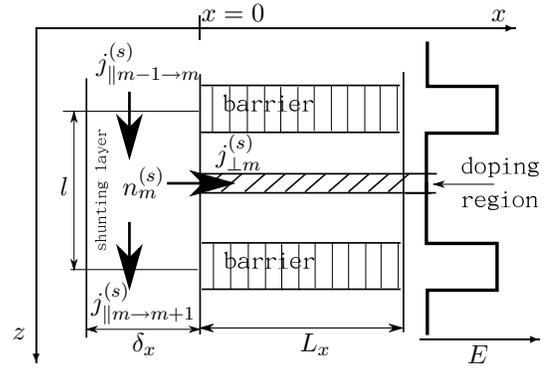}
\caption{Schematic of the shunted SL. The growth direction is along
the $z$ direction and the quantum wells are parallel to the $x$
direction. The SL is located at $x>0$ and the shunt is at $x<0$. The
thick line on the right is the potential energy of an electron in
the conduction band of the SL.} \label{model}
\end{figure}

\begin{figure}
\includegraphics[width=7cm]{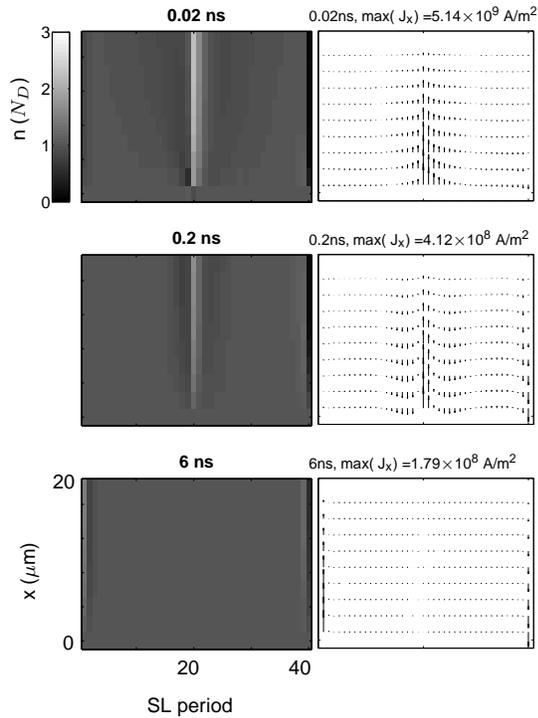}
\caption{Charge density plots (left column) and current vector plots
of $j_{\perp m}(x)$ (right column) for a SL with $L_x=20$~$\mu$m,
$U=2.1$~V and $\sigma=0.04$~$(\Omega$m$)^{-1}$ at 0.02, 0.2 and
6~ns. Initial condition is a CAL at the center of the SL. The shunt
is at the bottom. The color bar on the left of the first contour
plot is the scale encoding in units of $N_D$ used throughout the
paper.} \label{20um}
\end{figure}

\begin{figure}
\includegraphics[width=7cm]{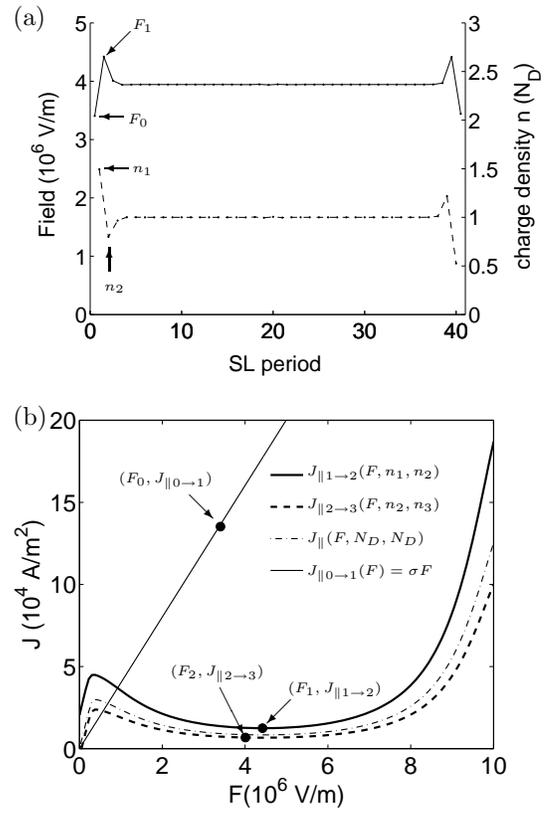}
\caption{The steady state for a SL with $L_x=20$~$\mu$m, $U=2.1$~V and $\sigma=0.04$~$(\Omega$m$)^{-1}$ at $x=L_x$:
(a) field profile (solid line) and charge density (dashed line), (b) The solid dots indicate the actual current
operation points on the local vertical current field characteristics $j_{\parallel m\rightarrow m+1}(F,n_{m},n_{m+1})$. }
\label{20um2}
\end{figure}

\begin{figure}
 \includegraphics[width=7cm]{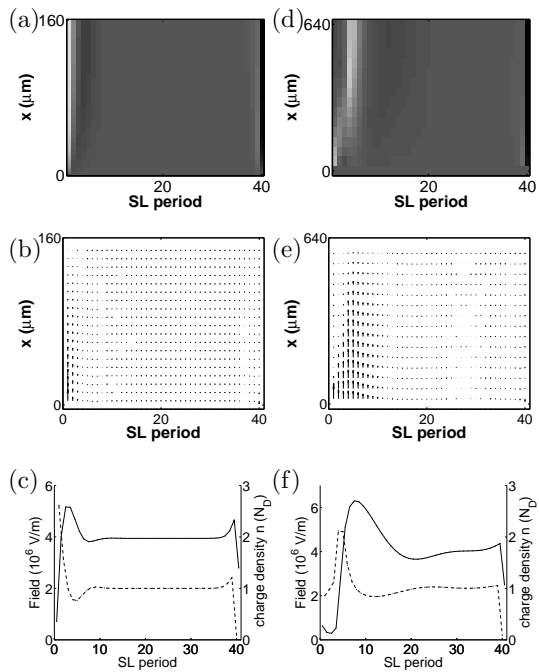}
\caption{Steady states: (a), (d) Charge density plots, (b), (e)
current vector plots and (c), (f) field profile (solid line) and
charge density (dashed line) at $x=L_x$ for $L_x=160$~$\mu$m (left
column) and $L_x=640$~$\mu$m (right column), respectively, with
$U=2.1$~V and $\sigma=0.04$~$(\Omega$m$)^{-1}$.} \label{640um}
\end{figure}

\begin{figure}
 \includegraphics[width=7cm]{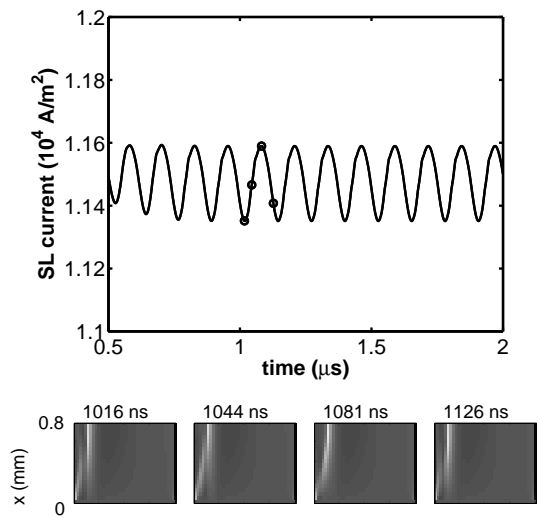}
 \caption{SL current density $J_{\text{SL}}/L_x$ and snapshots of charge density distribution for $L_x=0.8$~mm,
 $U=2.1$~V and $\sigma=0.04$~$(\Omega$m$)^{-1}$. The times of the snapshots are marked as solid circles in the upper panel.}
\label{0.8mm}
\end{figure}

\begin{figure}
\includegraphics[width=7cm]{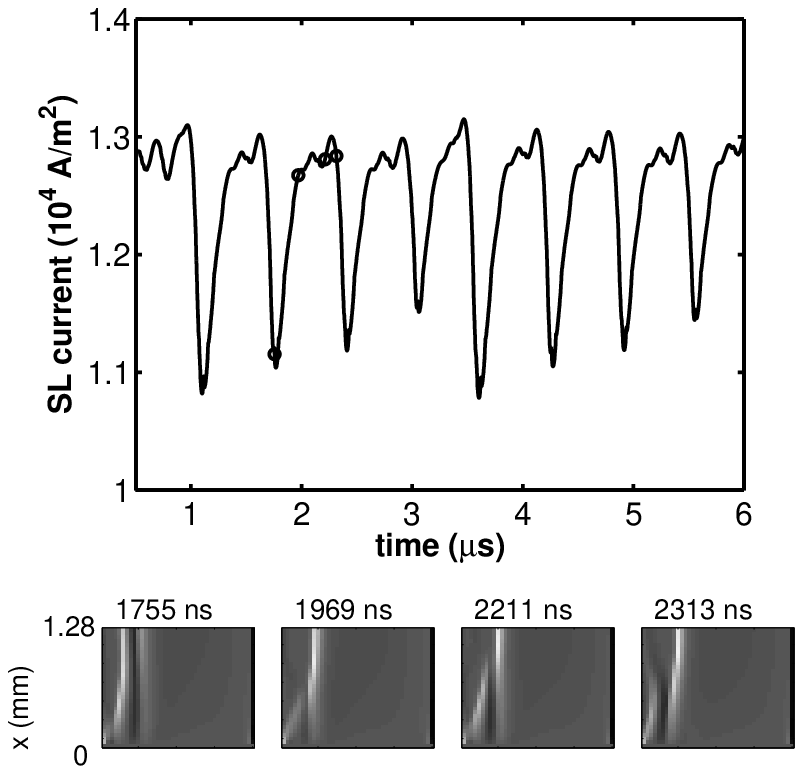}
\caption{Same as Fig.~\ref{0.8mm}, but with $L_x=1.28$~mm.}
\label{1.28mm}
\end{figure}

However, no such devices have been realized to date more than three
decades after their proposal because the NDC causes space-charge
instability. Although Bloch oscillations have been observed
experimentally in undoped SLs \cite{FeldmannL92} by studying optical
dephasing of Wannier-Stark ladder\cite{MendezA88} excitations using
degenerate four-wave mixing,\cite{YajimaT79, LeoS91} the gain
achievable is too small to build electrically active Bloch
oscillators. For high current densities, the space-charge
instability causes moving charge accumulation layers (CALs) and
charge depletion layers (CDLs) and thus the SL exhibits oscillations
similar to the Gunn effect.\cite{Gunn63} While devices based on
these oscillations may operate in the microwave range, they do not
extend to the THz region.\cite{SchomburgH97}

The lack of suitable THz radiation sources and detectors hampers the
technological exploitation of the frequency regime spanning from 300
GHz to 10 THz. Quantum cascade laser devices have been shown to
operate in the THz range for temperatures up to 164
K.\cite{WilliamsK05} On the other hand, if superlattice-based Bloch
oscillators could be successfully realized they might be expected to
have certain advantages relative to the quantum cascade
structures.\cite{WillenbergD03} Recently, rapid progress in THz
technology \cite{THz} including biomedical sensing,
three-dimensional imaging and chemical agent detection has attracted
renewed attention to Bloch oscillators. Some structures have been
proposed to stabilize the field in the SL against NDC-related
instabilities. One scheme theoretically proposed by Hyart \textit{et
al.}\cite{HyartA08} is the dc-ac-driven SL which requires the
presence of an initial THz pump. The SL is biased in the NDC region
under a DC electric field, initially superposed with an AC pump
electric field which stabilizes the field
distribution.\cite{Kroemer} Then the initial pump field can be
gradually turned off when THz oscillation has been already
established in the SL. Another suggestion is to stack a few short
SLs, where domains are not able to form.\cite{SavvidisL04} These
short SLs are separated by heavily doped material, and an increase
in terahertz transmission at dc bias has been observed.

Yet another scheme is to open a shunting channel parallel to the SL,
similar to a method that has been used to stabilize tunnel diode
circuits.\cite{BaoW06, WallmarkV63} Daniel \textit{et al.}
\cite{DanielG05} used a distributed nonlinear circuit model to
simulate the electric field domain suppression in a SL. They have
shown that the shunt is able to suppress the voltage inhomogeneity
above a critical bias voltage which depends on the shunt width, the
SL width, and the shunt resistivity. However, the circuit model does
not include aspects of the electronic tunneling transport that
appear to play an important role in SL behavior. The model possesses
only a global coupling since the elements are connected in series
and the $I-V$ characteristic of each element is fixed. On the other
hand, the SL model has a more complex structure that has both a
global coupling due to the applied voltage constraint as well as a
nearest neighbor coupling arising from the varying charge densities
that dynamically change the local current density vs. field ($J-F$)
characteristics. As a result, the nonlinear circuit model of Daniel
\textit{et al.} is not able to exhibit connected field domains or
current self-oscillations that are observed in SL structures both
theoretically and experimentally.\cite{BonillaG05}

In similar work by Feil \textit{et al.},\cite{FeilT05} a side
layer is grown on the cleaved edge of a lightly doped GaAs/AlGaAs SL, such that a 2D electron gas is
formed at the interface between the SL and the side layer. The lightly doped SL serves two purposes:
(i) to provide a modulated potential for the 2D electron gas at the interface so that under this
periodic potential, the electron gas becomes a \textit{surface SL} with one lateral dimension;
(ii) to provide a uniform field to
this surface SL since a lightly doped SL can maintain a uniform field
under external bias. While the suppression of field instabilities
has been reported in this type of SL, it is still not clear whether
this lateral structure will be useful as a THz oscillator.

In this paper, we study an extension of a well-established model of electronic transport in
weakly-coupled superlattices by adding a shunting side
layer. Our treatment includes the effect of lateral electronic
(i.e., horizontal) transport within each of the quantum well layers.
Here, the vertical electron dynamics is associated with sequential
resonant tunneling between weakly-coupled quantum well layers,
rather than miniband transport or Wannier-Stark hopping as occurs
for strongly-coupled SLs. Although the Bloch oscillator generally
requires a strongly-coupled SL, the weakly-coupled SL has similar
NDC features in I-V characteristics and similar current self-oscillations occur due to recycling of electronic fronts.\cite{WillenbergD03, Wacker02}

In the next section, we establish a two-dimensional model for
describing the current flow and dynamical electric field profile in
a shunted SL. In section III, we discuss the extremely different
time scales involved in this model, which are challenges to
numerically solving it. In section IV, we numerically explore the
effect of a high quality shunt on the dynamics of SLs as the lateral
size of the SL is varied, and show that the uniform field
configuration is stable, provided that the shunt and shunt
connection have high enough quality and the SL lateral extent is not
too great. In section V, we choose a laterally narrow SL and study
the dependence of the SL dynamics on the shunt properties. The
transition from a stable uniform field configuration to static field domains is found to be complex and the bifurcations involved in this transition are discussed. The Appendix presents details of the numerical methods employed.

\section{Laterally extended model of the superlattice with shunt layer}
\label{sec:later-extend-model} Weakly-coupled semiconductor superlattices have been successfully described by the sequential resonant tunneling model over the past several years.\cite{SCH01,Wacker02,BonillaG05, XuT07} However, previous works usually consider only the dynamics along the growth (vertical) direction of the SL and ignore the dynamics in the in-plane (lateral) direction, i.e., treat each period as an infinitely large plane with uniform charge density. More recently, Amann \textit{et al.}\cite{AmannS05} developed a theoretical framework which describes both lateral and vertical electronic dynamics. Here, we extend this framework to include the effects of a shunting side layer.

The structure of the shunted SL is shown in Fig.~\ref{model}. Each
quantum well forms a slab that is parallel to the $x-y$ plane, with
cross sectional dimensions $L_x$ and $L_y$. There are $N$ such
quantum wells stacked on top of each other in the $z$ direction,
sandwiched between an emitter layer and a collector layer. The shunt
layer is located between $-\delta_x\leq x\leq0$, with thickness
$\delta_x$. The SL period is $l=w+d$, where $w$ and $d$ are the width of the quantum well and width of the barrier, respectively. The external voltage is applied in the $z$ direction, across the emitter and the collector.

Inside the SL, the electrons are localized within one quantum well due to the relatively thick
quantum barriers. Furthermore, the electrons are assumed to be at local equilibrium and the
local \textit{two-dimensional} charge density at time $t$ is denoted by $n_m(x,y,t)$, where $m$
is the well index, $x$, $y$ are the in-plane coordinates. The charge continuity equation in the
SL can be written as:
\begin{equation}
e\; \dot{n}_m(x,y,t)=j_{\parallel m-1\rightarrow m}-j_{\parallel m\rightarrow m+1}-\nabla_\perp \cdot
\mathbf{j}_{\perp m},
\label{ndynamics}
\end{equation}
where
\begin{equation}
\nabla_\perp=\mathbf{e}_x\frac{\partial}{\partial x}+\mathbf{e}_y\frac{\partial}{\partial y},
\end{equation}
and $j_{\parallel m-1\rightarrow m}$ denotes the three dimensional
vertical current in $z$ direction tunneling through each barrier
(units: [A/m$^2$]) and $\mathbf{j}_{\perp m}$ is the lateral
two-dimensional current density (units: [A/m]). The electron charge
is $e<0$. The $y$-dependence is ignored and Eq.~(\ref{ndynamics}) can
be rewritten as:
\begin{equation}
e\; \dot{n}_m(x,t) =
j_{\parallel m-1\rightarrow m}\; -\,j_{\parallel m\rightarrow m+1}\;
 -\, \frac{\partial j_{\perp m}(x)}{\partial x}.
\label{SLdynamics}
\end{equation}

The local vertical tunneling current $j_{\parallel m\rightarrow
m+1}$ through each barrier is described by the sequential resonant
tunneling model which has been derived using different
methods;\cite{Wacker02, BonillaG05, XuT07} in this paper, we have
used the same form as in Refs. \onlinecite{BonillaG05, XuT07}. This
tunneling current depends on the electric field $F_{\parallel m}(x)$
across the barrier through which the tunneling occurs and the electron
charge densities $n_{m-1}(x)$ and $n_m(x)$ in the neighboring quantum
wells of this barrier. Thus, the tunneling current has the functional form:
\begin{equation}
j_{\parallel m-1\rightarrow m}(x)=\,j_{\parallel m-1\rightarrow m}[F_{\parallel m}(x),n_{m-1}(x), n_m(x)].
\end{equation}
The tunneling current densities through the emitter and collector layers are modeled by Ohmic boundary conditions,\cite{AmannS05} that is, $j_{\parallel 0\rightarrow 1}(x)=\sigma F_{\parallel 0}(x)$, and
$j_{\parallel N\rightarrow N+1}(x)=\sigma F_{\parallel N}(x)n_N/N_D
$, with contact conductivity $\sigma$ and two-dimensional doping
density $N_D$ in each well.

The lateral dynamics is caused by the in-plane current
$\mathbf{j}_{\perp m}$ which consists of a drift part and a
diffusion part. When the $y$-dependence is ignored, this becomes
\begin{equation}
j_{\perp m}(x)=-e\mu n_m F_{\perp m}-eD_0\frac{\partial n_{m}}{\partial x}
\end{equation}
where $F_{\perp m}(x)$ is the in-plane component of the electric
field at $x$ in well $m$, $\mu$ is the mobility and $D_0$ is the
diffusion coefficient. The generalized Einstein relation
\cite{CHE00} establishes the connection between $\mu$ and $D_0$ for
arbitrary two-dimensional electron densities including the
degenerate regime:
\begin{equation}
D_0(n_m)=\frac{n_m}{-e\rho_0(1-\mathrm{exp}[-n_m/(\rho_0k_BT)])}\mu
\end{equation}
with the two-dimensional density of states $\rho_0=m^*/(\pi \hbar^2)$, where $m^*$ is the electron effective mass. Here we assume that $\mu$ and $D_0$ are fixed.

Both the lateral and vertical currents depend on the electrical fields which in turn depend on the scalar
potential $\phi_m(x,y)$. The potential can be solved by the Poisson equation
\begin{equation}
\Delta \phi_m(x,y)=(\Delta_\perp+\Delta_\parallel)\phi_m(x,y)=-\frac{e}{l\epsilon_r\epsilon_0}(n_m-N_D),
\label{Poisson}
\end{equation}
with
\begin{eqnarray}
\Delta_\perp \phi_m(x)&=&\frac{\partial^2}{\partial x^2}\phi_m(x),\\
\Delta_\parallel \phi_m(x)&=&\frac{\phi_{m-1}(x)-2\phi_m(x)+\phi_{m+1}(x)}{l^2},
\end{eqnarray}
where $\epsilon_r$ and $\epsilon_0$ are the relative and absolute permittivity, respectively. Then the field can be calculated as
\begin{eqnarray}
F_{\parallel m}(x,y)&=&\frac{\phi_{m+1}(x)-\phi_m(x)}{l},\nonumber \\
F_{\perp m}(x)&=&-\frac{\partial \phi_m(x)}{\partial x}.
\label{SLfield}
\end{eqnarray}
Here we solve the Poisson equation using an approximation method assuming that
the typical structures in the lateral direction vary on a length scale much longer than the mean free path of the
degenerate electrons.\cite{AmannS05}

The drift-diffusion dynamics of the shunting layer is similar to
that of the lateral dynamics within each SL quantum well. First, we
neglect $x$-dependence in the shunt, that is, the shunt is collapsed
into a single layer along the $z$-direction. Note also that unlike
the SL, which possesses an intrinsic discreteness along $z$
direction, the shunt is a continuous layer. Therefore, we make a
further approximation that the shunt is divided into blocks aligned
with the periods of the SL and that the charge density is locally
uniform within each block. This assumption not only provides the
discretization required by numerical simulation, but also matches
the dynamics of the shunt with that of the SL. With these two
assumptions, we can write down the continuity equation in the $m$-th
shunt block as follows:
\begin{equation}
e\; \dot{\tilde{n}}^{(s)}_m(t)\cdot \delta_x\, l\, L_y = \,j^{(s)}_{\parallel m-1\rightarrow m}\cdot \delta_x\, L_y
\, - \,j^{(s)}_{\parallel m\rightarrow m+1} \cdot \delta_x\, L_y
\, - \, \tilde{j}^{(s)}_{\perp m} \cdot l \,L_y,
\label{shuntderive}
\end{equation}
where the superscript $(s)$ denotes the quantities in the shunt and the tilde denotes that the quantities are
three-dimensional, i.e.,
\begin{equation}
n^{(s)}_m = \tilde{n}^{(s)}_m\cdot l; \hspace{4mm}
j^{(s)}_{\perp m} = \tilde{j}^{(s)}_{\perp m} \cdot l.
\end{equation}
Here, the quantity $j^{(s)}_{\perp m}$ denotes the lateral current
that flows between the shunt and the SL through their interface.
Then we can write Eq.~(\ref{shuntderive}) in the form:
\begin{equation}
e\; \dot{n}^{(s)}_m(t)  =
\,j^{(s)}_{\parallel m-1\rightarrow m}
\, - \,j^{(s)}_{\parallel m\rightarrow m+1}
\, - \, \frac{ j^{(s)}_{\perp m} }{\delta_x},
\label{shuntdynamics}
\end{equation}

Note that the vertical current in the shunt has a very different
form than the tunneling current in the SL. It follows a similar
dynamics as the in-plane current in the SL quantum wells and is
related to the three-dimensional charge density in the shunt:
\begin{equation}
j^{(s)}_{\parallel m-1\rightarrow m}= -e\mu \tilde{n}^{(s)}_m F^{(s)}_{\parallel m}-e D_0
\frac{\partial \tilde{n}^{(s)}_m}{\partial z}.
\label{currentinshunt}
\end{equation}
Here we assume the mobility $\mu$ and the diffusion coefficient
$D_0$ have the same values as in the SL.

Next, we examine the lateral current that connects the shunt and the
quantum well layer within the  SL:
\begin{equation}
j^{(s)}_{\perp m}=-e\mu n_m(x=0) F_{\perp m}-D_0\nabla_\perp n_m
\bigg|_{ x=0^+}.\label{eq:2}
\label{currentconnect}
\end{equation}
In this equation, the boundary should be defined at $x=0^+$ for
calculation of both the current and the potential in the shunt.
Since the shunt is assumed to be uniform in $x$ direction, defining
the above equation at $x=0^-$ implies that $F_{\perp m}$ and
$\nabla_\perp n^{(s)}_m$ are zero which would lead to zero boundary
current. Another advantage of choosing the boundary at $x=0^+$ is that the potential in the shunt should be equal to the potential in the SL close to its boundary,
i.e., $\phi^{(s)}_m(x<0)=\phi_m(x=0^+)$, since the potential is continuous everywhere. This relation allows us to equate the potential in the shunt with that at the inner boundary of the SL. So the potential at the boundary of the solution of Eq.~(\ref{Poisson}) is just the potential in the shunt. The fields required to calculate the current in Eq.~(\ref{eq:2}) can be obtained by
\begin{eqnarray}
F^{(s)}_{\parallel m}(x)&=&\frac{\phi^{(s)}_{m+1}(x)-\phi^{(s)}_m(x)}{l},\nonumber\\
F_{\perp m}(0^+)&=&-\nabla_\perp\phi_m(x)\bigg|_{x=0^+}.
\label{shuntfield}
\end{eqnarray}
The charge density and its normal gradient at the boundary are
\begin{eqnarray}
n_m(x=0)&=&\frac{n_m(0^+)+n^{(s)}_m(0^-)}{2}, \\
\nabla_\perp n_m\bigg|_{x=0^+}&=&\lim_{\Delta x \to  0^+}\frac{n_m(\Delta x)-n^{(s)}_m}{\Delta x}.
\label{eq:1}
\end{eqnarray}
Here we also note the possible effects of energy band structure of
the shunted SL and the doping density in the shunt. In the above
discussion, the situation has been simplified because no band
bending is included. However, variations in doping densities in the
shunt and the SL can cause band bending effects at the interface.
Even if the shunt is doped to have the same Fermi level as that in
the SL so that little band bending might be expected, there are
other issues that impact the connection quality between the shunt
and the SL, for example, the presence of trap states or a thin oxide layer. To quantify the quality of the connection between the SL and the shunt, we introduce a
parameter $0 \le a \le 1$ such that $a=1$ corresponds to a perfect connection and $a=0$ corresponds to no connection. We modify Eq.~(\ref{currentconnect}) to be
\begin{equation}
j^{(s)}_{\perp m}=a\cdot\left(-e\mu n_m(x=0) F_{\perp m}-D_0\nabla_\perp n_m \bigg|_{ x=0^+}\right).
\end{equation}
The relationship between specific values of parameter $a$ and microscopic models of conduction across the shunt-SL interface are discussed elsewhere.\cite{Xuu}

Similarly, we introduce a separate parameter $b > 0$ that allows us
to model the effect of having different doping density and/or mobility in the shunt vs. SL quantum wells. Also recognize that the field in the shunt is almost uniform and $n^{(s)}_m\approx N_D^{(s)}$ when the conductance in the shunt is high, where $N_D^{(s)}$ is the doping density in the shunt. This leads to the following modification of Eq.~(\ref{currentinshunt}),
\begin{equation}
j^{(s)}_{\parallel m-1\rightarrow m}= -e\mu \tilde{n}^{(s)}_m F^{(s)}_{\parallel m}-e D_0
\frac{\partial \tilde{n}^{(s)}_m}{\partial z} \approx
-e b \mu^{(s)} \tilde{N}_D F^{(s)}_{\parallel m},
\end{equation}
where $b \mu \tilde{N}_D = \mu^{(s)} \tilde{N}^{(s)}_D$. Note that $b>1$ when the doping density in the shunt is greater than that in the quantum wells and $b$ is much less than one when the shunt is weakly conducting so that only a small fraction of the total vertical current flows through it.

It is also useful to point out that the total current,
\begin{equation}
J=\left(\epsilon_r \epsilon_0 \dot{F}^{(s)}_{\parallel m}+{j}^{(s)}_{\parallel m\rightarrow m+1}\right)\cdot\delta_x +
\int_0^{L_x}\left(\epsilon_r \epsilon_0\dot{F}_{\parallel m}+j_{\parallel m\rightarrow m+1}\right)dx,
\end{equation}
is the same for each period.  To show this, note that the Poisson
equation can be written as
\begin{equation}
\nabla\cdot(\mathbf{F}_\perp+\mathbf{F}_\parallel)=\frac{e}{l
\epsilon_r \epsilon_0}(n_m-N_D),
\end{equation}
or
\begin{equation}
\frac{F_{\parallel m}-F_{\parallel m-1}}{l}+\frac{\partial F_\perp}{\partial x}=\frac{e}{l \epsilon_r \epsilon_0}(n_m-N_D).
\end{equation}
\begin{widetext}
Substituting the above equation into Eq.~(\ref{SLdynamics}) yields
\begin{equation}
l \epsilon_r \epsilon_0 \frac{d}{dt}\left(\frac{F_{\parallel m}-F_{\parallel m-1}}{l}+\frac{
\partial F_\perp}{\partial x}\right) = j_{\parallel m-1\rightarrow m}\; -\,j_{\parallel m\rightarrow m+1}\; -\,
\frac{\partial j_{\perp m}(x)}{\partial x}.
\end{equation}
Then, one integrates both sides of the preceeding equation with
respect to $x$ from $-\delta_x$ to $L_x$. Due to the vanishing
boundary conditions $F_\perp(-\delta_x)=F_\perp(L_x)=0$ and
$j_{\perp m}(-\delta_x)=j_{\perp m}(L_x)=0$, the lateral terms in
the above equation integrate to zero. This yields
\begin{equation}
\epsilon_r \epsilon_0 \frac{d}{dt}\int_{-\delta_x}^{L_x}F_{\parallel m}dx +\int_{-\delta_x}^{L_x}\,
j_{\parallel m\rightarrow m+1} dx = \epsilon_r \epsilon_0 \frac{d}{dt}\int_{-\delta_x}^{L_x}F_{\parallel m-1}dx +
\int_{-\delta_x}^{L_x}\,j_{\parallel m-1\rightarrow m} dx.
\end{equation}
\end{widetext}
which shows that the total current is independent of the well index
$m$. Note that the current through the shunt will be the dominating
contribution to the total current of a SL if the shunt is thick and
well-conducting. Even a completely disconnected shunt (i.e. $a=0$)
contributes a constant current of $J^{(s)}_0=\delta_x e \mu N_D
U/(Nl+d)$ to the total current $J$ of a homogeneous SL. Since we are
interested in effects arising from the interaction between  the SL
and the shunt, we will in the following discuss the current dynamics
on the basis of the SL current defined by
$J_{\text{SL}}(t)=J(t)-J^{(s)}_0$.

\section{Parameters and Time scales}
The parameters that we use in the simulation are listed in Table I.
\begin{table}[ht]
\caption{Parameters used for the shunted SL.}
\begin{ruledtabular}
\begin{tabular}{cccccccc}
$N$ & $N_D$ & $w$ & $d$ & $\mu$ & $D_0$ & $T$ & $\epsilon_r$\\
- & (m$^{-2}$) & (nm) & (nm) & (m$^2$/Vs) & (m$^2$/s) & (K) & -\\\hline
40 & $1.5\times10^{15}$ & 9 & 4 & 10  & 0.015 & 5 & 13.18\\
\end{tabular}
\end{ruledtabular}
\label{typicalvalues}
\end{table}
 We found that there are very different time scales in this complex structure which requires an implicit
 method of numerical iteration. The first time scale $\tau_b$ is the dielectric relaxation time in the
 bulk material both in the shunt and in each quantum well in the SL. It is determined by the doping density.
 We know that the conductivity $g$ is proportional to the charge density
\begin{equation}
g\approx e\mu N_D/l \sim 1.6\times 10^{-19}\times 10\times 10^{23}\mathrm{(\Omega m)}^{-1}\sim 10^5 \mathrm{(\Omega m)}^{-1}
\end{equation}
So the dielectric relaxation time in the shunt layer and within each
quantum well is approximated as
\begin{equation}
\tau_b=\frac{\epsilon_r \epsilon_0}{g}\sim
\frac{0.1\times10^{-9}}{10^5} \mathrm{(s)}\sim 10^{-15}
\mathrm{(s)}
\end{equation}
which is relatively fast due to the high conductivity. This is the
time it takes for a fluctuation in the charge density to be
neutralized within either the shunt layer or quantum wells.

The second time scale $\tau_t$ is the one in the vertical dynamics.
According to the sequential resonant tunneling model, the vertical
current is to the order of $10^{-4}$~(A/m$^2$) and the positive
differential conductivity $g_t$ is of order $0.1$~($\Omega$
m)$^{-1}$. Thus, $\tau_t = \epsilon_r\epsilon_0/g_t \sim 10^{-9}$~s,
a much larger time scale than $\tau_b$. Moreover, from numerous
previous works, we also know that the behavior of the electrons in
the vertical direction is not simply dielectric relaxation. More
complex phenomena, such as current self-oscillation, or injected
dipole relocation due to switching, have much longer time scales
ranging up to microseconds. The time scale $\tau_t$ sets a lower
limit of the time scales for these nonlinear processes.

Another important time scale $\tau_i$ is the time that it takes to
carry away or supply the electrons in the SL through the shunt.
Because the vertical processes are relatively slow, if the shunt has
good connection and high conductance, the electrons will move
laterally, pass through the intersection between the quantum well
and the shunt, and drift away through the shunt. This time scale
$\tau_i$ is considerably larger than $\tau_b$ since the electrons
have to move into the shunt first.  Later we will see that it takes
1 ns to deplete a full CAL in a small SL.  The presence of extremely
different time scales means that the numerical integration is a
stiff problem and this suggests the use of an implicit method. The
numerical procedure is described in the Appendix.

\begin{figure}
\includegraphics[width=7cm]{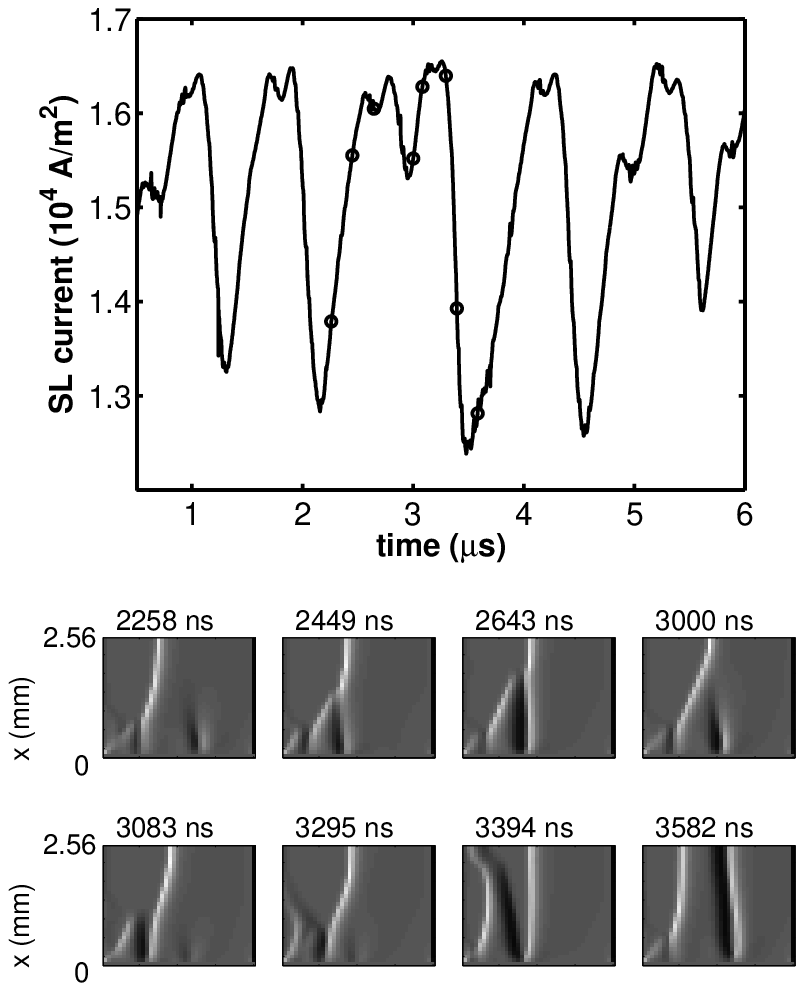}
\caption{Same as Fig.~\ref{0.8mm}, but with $L_x=2.56$~mm.}
\label{2.56mm}
\end{figure}

\begin{figure}
\includegraphics[width=7cm]{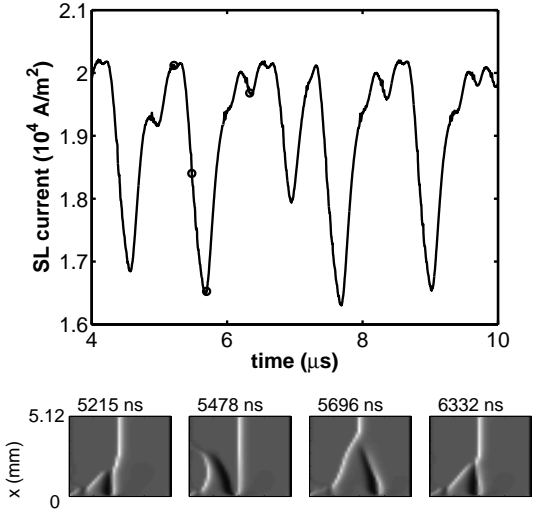}
\caption{Same as Fig.~\ref{0.8mm}, but with $L_x=5.12$~mm.}
\label{5.12mm}
\end{figure}

\begin{figure}
 \includegraphics[width=6cm]{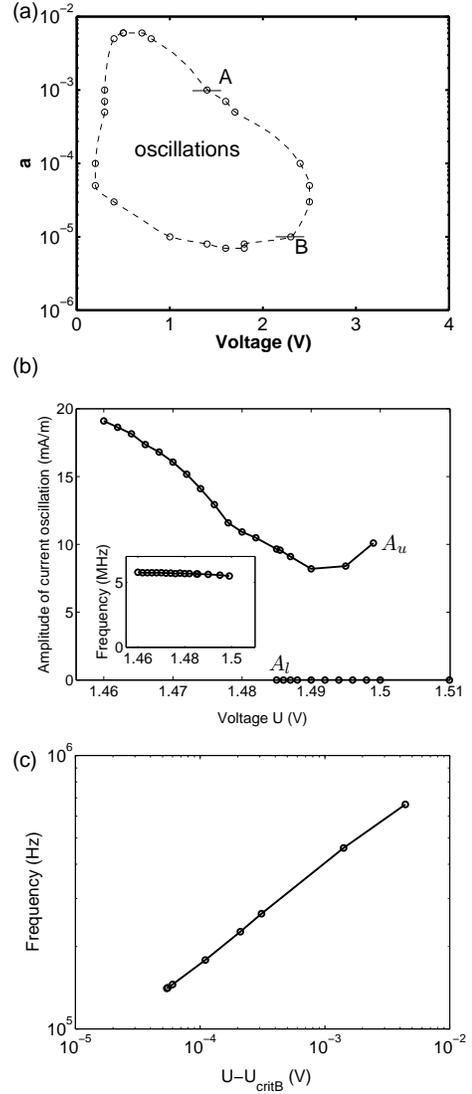}
\caption{(a) Bifurcation diagram for $\sigma=0.04$~$(\Omega$m$)^{-1}$, $L_x = 20$~$\mu$m, $b=1.00$.
Dashed curve shows the approximate boundary of the oscillatory region and location of studied bifurcation
points $A$ and $B$. (b) Bifurcation scenario at $A$ for $a = 1.00\times 10^{-3}$: amplitude vs. voltage
(main figure) and frequency vs. voltage (inset). Points $A_u$ and $A_l$ denote the endpoints of the upper
and the lower branches, respectively. (c) Bifurcation scenario at $B$ for $a = 1.00\times 10^{-5}$:
scaling of frequency vs. voltage (double logarithmic plot); $U_{critB}=2.30441$~V.}
\label{bifurAB}
\end{figure}

\begin{figure}
\includegraphics[width=7cm]{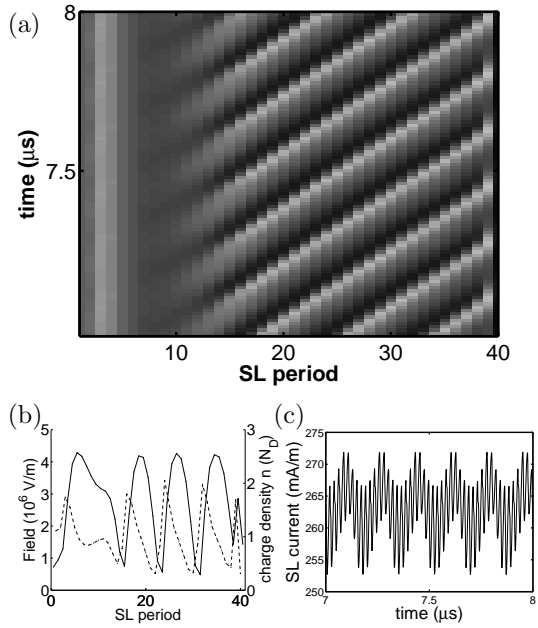}
\caption{(a) Charge density distribution evolving in time, (b) a snapshot of field profile (solid line)
and charge density profile (dashed line) at $t=8$~ns and (c) SL current $J_{\text{SL}}$ on the upper
branch of Fig.~\ref{bifurAB}b. Parameters: $a = 1.00\times 10^{-3}$, $U = 1.46$~V, $\sigma=0.04$~$(\Omega$m$)^{-1}$.}
\label{osciA}
\end{figure}

\begin{figure}
\includegraphics[width=7cm]{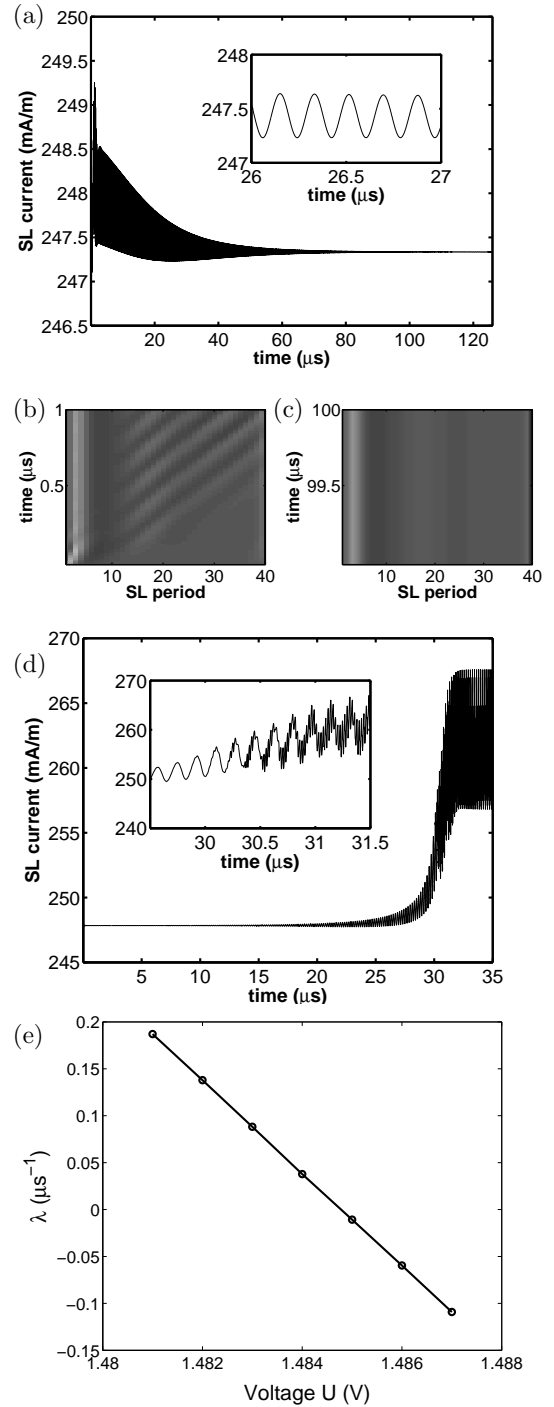}
\caption{(a) SL current $J_{\text{SL}}$ vs. time for $U=1.486$~V (the inset shows an enlargement),
(b), (c) charge density distributions evolving for two different time intervals for voltage near bifurcation point
$A_l$. (d) SL current for $U=1.481$~V. (e) The rate $\lambda$ of exponential decay $\lambda<0$ (or increase $\lambda>0$)
of oscillation amplitude versus applied voltage $U$. Parameters: $a = 1.00\times 10^{-3}$,
$\sigma=0.04$~$(\Omega$m$)^{-1}$.}
\label{subHopf}
\end{figure}

\begin{figure}
\includegraphics[width=7cm]{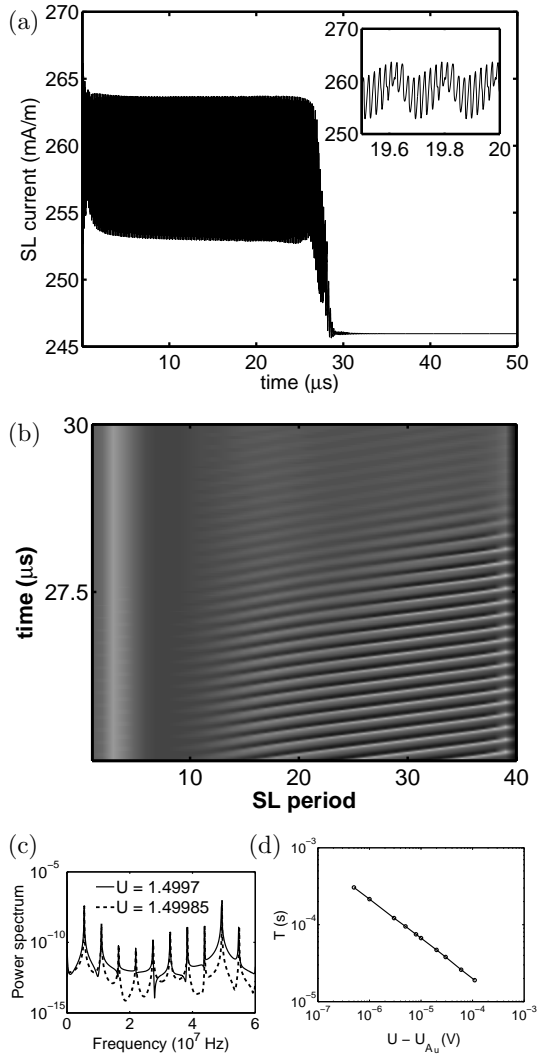}
\caption{(a) SL current $J_{\text{SL}}$ (the inset shows an enlargement), and (b) charge density distribution
evolving in time for $U = 1.49985$~V with $a = 1.00\times 10^{-3}$, $\sigma=0.04$~$(\Omega$m$)^{-1}$. (c)
Power spectrum data for oscillations at $U=1.4997$~V and $U=1.49985$~V. (d) The time $T$ for which the system
exhibits transient oscillations versus the applied voltage $U -U_{A_u}$. $U_{A_u}=1.499791$~V.}
\label{saddlenode}
\end{figure}

\begin{figure}
\includegraphics[width=7cm]{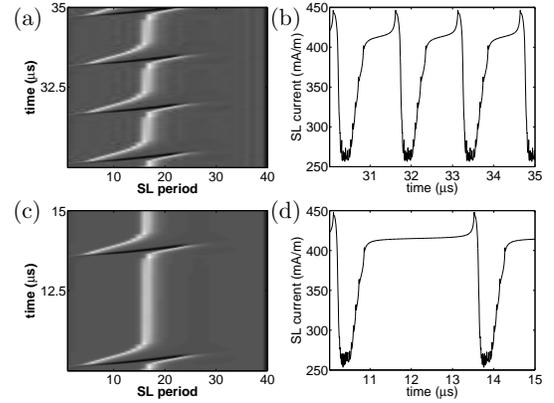}
\caption{Bifurcation scenario at point $B$ of Fig.9: (a), (c) charge density distributions vs. time, (b), (d) SL
current $J_{\text{SL}}$ for $U = 2.3$~V (upper panel) and $U=2.304$~V (lower panel), respectively, with
$a = 1.00\times 10^{-5}$, $\sigma=0.04$~$(\Omega$m$)^{-1}$.}
\label{bifurB}
\end{figure}

\begin{figure}
\includegraphics[width=7cm]{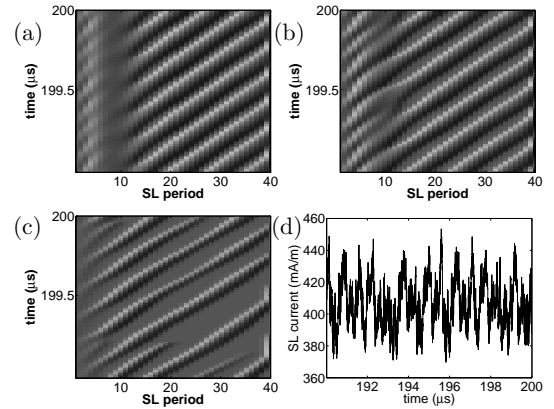}
\caption{Charge density distribution vs. time for $a = 1.00\times
10^{-3}$, $\sigma=0.04$~$(\Omega$m$)^{-1}$ at (a) $U = 1.2$~V, (b)
$U = 1$~V, and (c) $U = 0.5$~V. (d) SL current at $U = 0.5$~V.}
\label{chaos}
\end{figure}

\begin{figure}
\includegraphics[width=7cm]{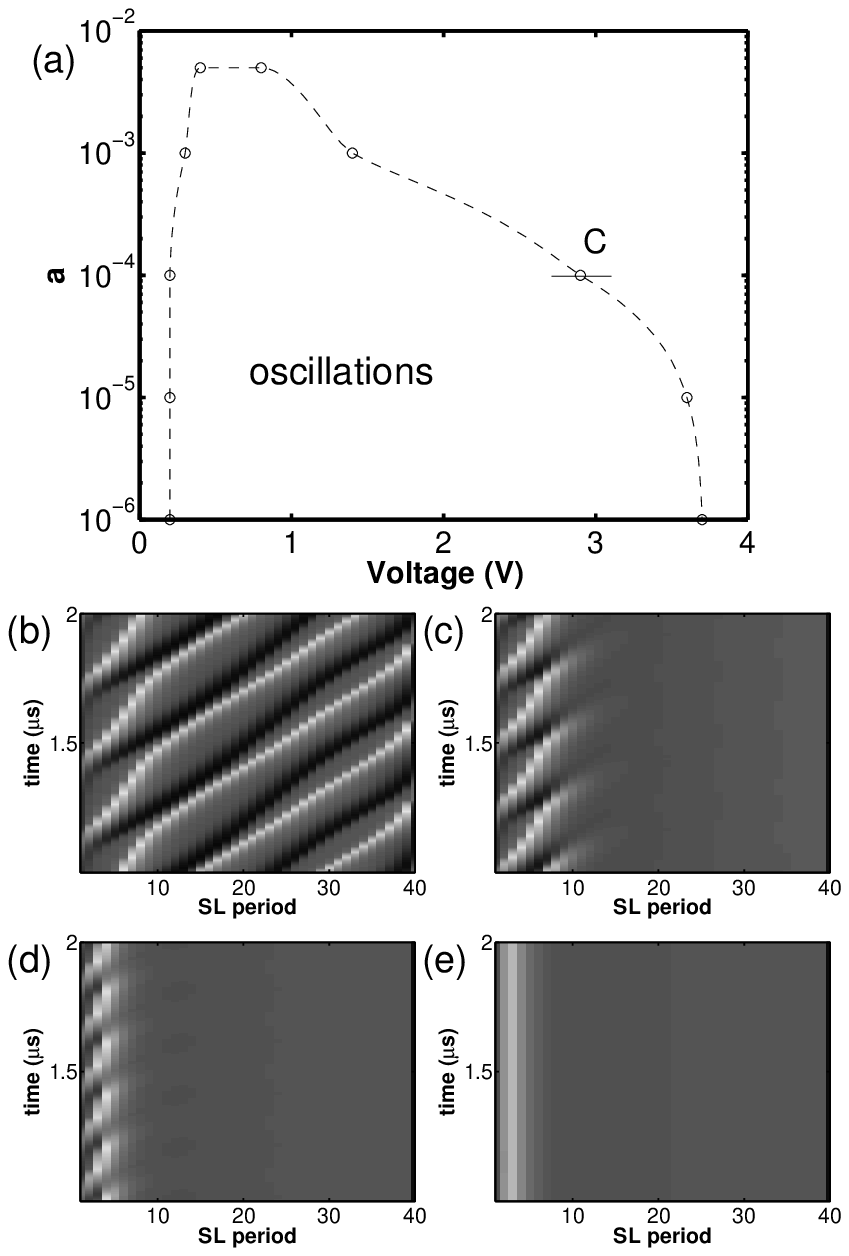}
\caption{(a) Bifurcation diagram for $\sigma=0.016$~$(\Omega$m$)^{-1}$. Charge density distribution vs. time
near bifurcation point $C$, $a = 1.00\times 10^{-4}$ at (b) $U=1.6$~V, (c) $U = 2.1$~V, (d) $U = 2.7$~V, and
(e) $U = 3.0$~V.}
\label{bifurC}
\end{figure}

\section{Dependence of shunting dynamics on the lateral size of the Superlattice}
In this section, we discuss the effects of the lateral size $L_x$ of
the SL with a high quality shunting layer, i.e., $a=b=1$. The shunting layer has a width $\delta_x$ such that
varying $\delta_x$ does not
affect the dynamics in the shunt. This is numerically confirmed even
for the chaotic case that we will discuss below, where a 80~nm
shunting layer has the same effect as a 8~mm one. This is because
$\tau_b$ is much smaller than $\tau_i$ and the electrons entering the
shunt are carried away so fast that a change in the shunt conductance does not change $\tau_i$. We will study the SLs with a relatively high contact conductivity $\sigma = 0.04$~$(\Omega$m$)^{-1}$. At this value of $\sigma$, without a shunt, the SL has a static high field domain near the emitter and a static low field domain near the collector separated by a static charge accumulation layer (CAL). Due to the high quality shunt the total current is dominated by the  contribution of the current through the shunt. As discussed at the end of Section \ref{sec:later-extend-model}, we will therefore consider the SL current $J_{\text{SL}}$. Also, since we are varying $L_x$, we scale current to current density.

\subsection{High quality shunting layer with small $L_x$}

Figure~\ref{20um} shows charge and current density plots for a
relatively narrow SL with lateral extent $L_x =20$~$\mu$m. The initial
state is prepared as a charge configuration for the SL without shunt
at total applied voltage $U=2.1$~V and shows a static charge
accumulation layer at the 20th period. After an interval of about 1~ns, the space charge
configuration is almost uniform. The in-plane current is plotted as
a vector field and shows the electrons in the CAL move in the
lateral direction (the opposite direction of the current) into the
shunt. We can see that when the system reaches steady state, the net
charge is almost neutral, i.e., $n=N_D$, everywhere in the SL and
the shunt. There are still some small lateral current flows at the first and the last period.

If we take a close look at the steady state, we find that there is a
small CAL at the first period and a CDL nearby (Fig.~\ref{20um2}(a)).
The situation is almost inverted near the collector. To better
understand this, we focus on the operation points near the emitter
shown in Fig.~\ref{20um2}(b) at $x = 20$~$\mu$m. In this case, the field is almost uniform in the SL and
each period is biased in the NDC region. The field across the first
barrier between the emitter and the first well will also have this
same value in the absence of charge accumulation in the first well.
This causes a vertical current from the emitter to the first period
(thin solid line in Fig.~\ref{20um2}(b))  which is much larger than
the vertical current in the corresponding NDC region of the SL.
Close to the shunt this extra current will give rise to a lateral
current which will quickly reach the shunt and is carried away by
the shunt. A little further away from the shunt where the lateral
current is not sufficient to completely neutralize this extra
current, a small CAL is formed in the first well which lowers the electric field and therefore the current across the first
barrier. At the same time, the electric field in the second barrier
is pushed above the uniform field, causing a very small CDL next
to the CAL. Similar arguments can be applied to the collector to
explain the appearance of a small CDL in the last quantum well.  The
overall effect is that a nearly uniform vertical electric field
configuration is stabilized for these conditions.

\subsection{High quality shunting layer with large $L_x$}
As the lateral size $L_x$ of the SL becomes larger, the CAL and CDL
near the emitter become more prominent (cf. Fig.~\ref{640um}(a)-(c),
$L_x =160$~$\mu$m) since with increasing distance to the shunt the
lateral current becomes less efficient at carrying away the excess
current from the emitter to the shunt.

For wider SL (cf. Fig.~\ref{640um}(d)-(f), $L_x =640$~ $\mu$m), the
field closer to the shunt is more uniform and the CAL is still
attached to the emitter. However, away from the shunt, the CAL
detaches from the emitter and locates itself in the first few
periods and the nonuniform field region becomes larger. This
behavior is due to the lateral current being insufficient to carry
away the extra current from the emitter. Thus, the CAL grows bigger
and tends to move toward the collector. With the center of the CAL
located in different wells at different $x$ positions, the lateral
gradients can be increased and a sufficient lateral current can be
sustained. The field profile at $x=640$~$\mu$m is plotted in
Fig.~\ref{640um}(f). Field domains are forming as the field is low
to the left of the CAL and high to the right of the depletion
region. In this case, the upstream CAL (closer to the emitter, at
the left bottom corner of Fig.~\ref{640um}(d)) and the downstream
CAL (closer to the collector, the wider one in Fig.~\ref{640um}(d))
are still connected and this is a time-independent steady state.

In the above case, the lateral size of the SL is just below a characteristic value for which the steady state loses stability to oscillatory behavior. Figure~\ref{0.8mm} ($L_x =800$~$\mu$m) shows the simulations of a slightly wider SL than considered above.  The large downstream CAL still stays in that position. However, due to the large size of the SL, the lateral current is not able to sustain a connected stable CAL. The small upstream CAL touches and breaks off from the downstream CAL periodically. There is a small amplitude oscillation in the total current which is shown in the top panel.

For an even wider SL (Fig.~\ref{1.28mm} with $L_x =$ 1.28~mm), the
upstream and downstream CALs are mostly disconnected. The upstream
CAL extends laterally into the SL and moves toward the downstream
CAL, (at time 1.969 ms). For certain times during the dynamical
evolution (not shown in Fig.~\ref{1.28mm}), the upstream CAL breaks
off from the emitter and reaches and merges with the downstream CAL.
Mostly, there is a depletion region forming between the upstream and
downstream CALs (2.211 ms).  This depletion region grow and dies
away as it merges with the upper CAL (1.969 ms). For certain times,
it grows into a full CDL extending throughout the structure and, in
this case, the upstream CAL also grows into a full CAL (1.755 ms).
Then all three fronts move downstream. The old downstream CAL and
the CDL quickly disappear and the new CAL formed by the upstream CAL
replaces the old CAL and stays at its position. Although these
behaviors are quite complicated, they are still periodic and during
each period, the upstream and downstream CALs merge several times.

However, for an extremely wide SL (Fig.~\ref{2.56mm}, $L_x=2.56$~mm), the behavior is apparently chaotic. The
effect of the shunt is to cause a CAL attached to the emitter near the shunt. For large values of $x$, the shunt
has less effect and this CAL detaches from the emitter, tends to move downstream to the collector and thus extends
toward the downstream CAL. Due to the large lateral size of the SL, the impact of the shunt layer becomes very weak
on the opposite side of the SL. Thus, the downstream CAL is located very close to the 20th period where it would be
in the absence of a shunting layer. The merging of the CALs described in last paragraph also appears here except
that the merging events are now difficult to predict and manifestly not periodic. Figure~\ref{5.12mm} shows the
behavior of a SL with $L_x=5.12$~mm. It should be noted that real SL samples rarely have such a large size. In
this case, the unstable dynamics only occurs in the portion of the SL closest to the shunt. In the portion of the
SL away from the shunt, a CAL is located at the 20th well, where the shunt has no apparent influence. Over time,
the lateral extension of this CAL changes.
When a large CDL collides with it at 5.696~ms, the static CAL shrinks
to a small size, causing a large dip in the current trace.  The
presence of such charge tripole configurations \cite{AMA02} of one
CDL and two CALs has already been shown to be associated with
chaotic behavior in one-dimensional SL models without lateral
dynamics.\cite{AMA02a}

To summarize, we are able to identify three characteristic length scales in the $x$ direction. The shortest one is the decay length $\bar L_x$ (of order 10~$\mu$m) at which the charge density in the first quantum well increases from $N_D$ at the SL-shunt interface to its maximum value (cf. Fig. \ref{640um}(a)-(c)). The next length scale (of order 200~$\mu$m) is the range above which the vertical field configuration loses uniformity and static field domains start to form (cf. Fig. \ref{640um}(d)-(f)). The longest length scale (of order 700~$\mu$m) is the width of the SL above which the steady state loses stability to oscillatory behavior. This implies that lateral uniformity in the electric field distribution can be expected when $L_x$ is smaller than the intermediate characteristic length scale. The shortest decay length $\bar L_x$ can be estimated by noting that the extra current coming from the emitter must be directed to the shunt by the negative gradient of the lateral current $J_\perp$, i.e., $\frac{\partial J_\perp(x)}{\partial x} = J_{\parallel 0\to 1} (x)-J_{\parallel 1\to 2}(x)<0$. Then there is approximately a decay length $\bar L_x$, at which the quantities such as $J_\perp(x)$, $n(x)$ and $F_x(x)$ approach asymptotic values exponentially. Calculation shows that $\bar L_x$ is of order $10$~$\mu$m for the parameters used in Table~I,\cite{Xuu} in agreement with our numerical results.

\section{Dependence of dynamical behavior on the shunt properties}
In the previous section, we have seen that the width of the SL determines the lateral dynamics of electronic transport and that the shunt can stabilize a nearly \textit{uniform} field configuration in sufficiently narrow SLs. Now we investigate the effects of the shunt properties on a small SL with width of 20 $\mu$m where the lateral field and electron density profiles are almost uniform. Since the charge density is almost uniform laterally, we modify the model such that \textit{the SL is collapsed to one point in $x$ direction}. This modification significantly reduces the complexity of the simulation. We first study the effects of connectivity parameter $a$ on a SL with conductivity $\sigma=0.04(\Omega$m$)^{-1}$ chosen as in the previous section. Then we study the effects of $a$ on a SL with lower contact conductivity $\sigma=0.016(\Omega$m$)^{-1}$, which corresponds to moving fronts and current self-oscillations in unshunted SLs,\cite{XuT07} and briefly discuss the effects of shunt conductivity parameter $b$ and width $\delta_x$. Since $L_x$ is fixed, we plot the unscaled SL current $J_{\text{SL}}$.

\subsection{Dynamical behavior vs. connectivity parameter $a$ for large contact conductivity}

Figure~\ref{bifurAB}(a) shows a bifurcation diagram using as the bifurcation parameters the connectivity parameter $a$ and the voltage $U$ for $\sigma=0.04(\Omega$m$)^{-1}$. There is a bounded region where the system exhibits periodic or chaotic oscillations, shown as the region enclosed by dashed lines in Fig.~\ref{bifurAB}(a). The value of the connectivity parameter $a$ of the oscillatory region ranges from about $6\times 10^{-3}$ to $7\times 10^{-6}$. In real samples, such a weak connection between the SL and the shunt could be associated with a potential barrier formed between the SL and the shunt due to band bending or an oxide layer.

For $a \gtrsim 6\times 10^{-3}$, the charge density in the SL is almost uniform except for a small CDL near the emitter, the same situation shown in Fig.~\ref{20um}. With the increase of voltage, this CDL becomes more prominent and there is an CAL in the first period. However, this CAL never detaches from the emitter for any value of voltage when $a\gtrsim 6\times10^{-3}$. This is reasonable because for $a\gtrsim 6\times10^{-3}$, the connection is strong enough that the shunt is able to maintain the field in the SL almost uniform.

Another stable region is $a\lesssim 7\times10^{-6}$. In this region, a static CAL is formed in the SL and located close to the position where it is expected when there is no shunt. This is also easy to understand because the connection is so weak that the shunt has almost no influence on the SL.

Between these two values of $a$, we have a transition region where oscillations occur for certain ranges of applied voltage. Here, the bifurcation scenarios by varying voltage are investigated for two sets of values ($A$ and $B$) of the control parameters.

The bifurcation for point $A$ occurs at $a=1.00\times10^{-3}$ and $U=1.485$~V (Fig.~\ref{bifurAB}(b), \ref{osciA}-\ref{saddlenode}). Inside the oscillatory region (approximately $U\leq1.485$~V), the charge density distribution in the SL oscillates and the oscillation only involves part of the SL (cf. Fig.~\ref{osciA}). There is a static CAL near the emitter but this is clearly detached from the emitter. The oscillation occurs in the wide region to the right of the CAL in the form of moving charge dipoles (CALs and CDLs), cf. Fig.~\ref{osciA}(a). However, at any given time, there are three to four pairs of dipoles present. From Fig.~\ref{osciA}(b), we can see that the charge densities have large amplitude fluctuations along the $z$ direction and the higher frequency component of the current oscillation is due to the movement of these dipoles (Fig.~\ref{osciA}(c)). This higher frequency $f_1$ is nine times the lower one $f_2$ at which the collector receives the moving dipoles. Here we observe the \textit{coexistence} of static CAL and steady moving fronts.

The bifurcation scenario of A is illustrated by Fig.~\ref{bifurAB}(b), where the amplitude of the current oscillation is plotted versus the applied voltage. There is a bistability region between $U \approx 1.485$~V and 1.50~V, where the system either oscillates (upper branch) or is in a steady state (lower branch).

The bifurcation at point $A_l$ at the end of the lower branch is studied in Fig.~\ref{subHopf}. When the system starts from a uniform configuration at $U = 1.486$~V, shown in Fig.~\ref{subHopf}(a), (b), (c), it first oscillates similar to the full oscillation in Fig.~\ref{osciA}, except that the CALs and CDLs are much smaller in Fig.~\ref{subHopf}(b). The oscillation gradually decays to a steady state where there is only a single stable CAL and no charge fronts to its right, as shown in Fig.~\ref{subHopf}(c). The amplitude of the current oscillation is quite small and decays to zero. The well-to-well hopping of the small charge fronts does not have an appreciable effect on the current oscillation form as found for the mature fronts in Fig.~\ref{osciA}(c). Instead, the shape of the current oscillation is smooth and sinusoidal and possesses a well-defined frequency. After a transient interval, the amplitude $A(t)$ of the current oscillation decays exponentially, i.e., $A(t)=A(t_0)\exp{\lambda t}$ and the rate $\lambda$ can be determined by fitting. It also should be mentioned that the initial state corresponding to the uniform field configuration falls into the basin of attraction of the upper oscillatory branch for $U\lesssim 1.486$~V. Hence, to obtain $\lambda$ for the lower branch, we start the system from the steady state of $U = 1.486$~V. This initial state is used for all the points of the lower branch. In the case of $U=1.481$~V, shown in Fig.~\ref{subHopf}(d), the amplitude of the current oscillation increases exponentially at first and after passing a certain threshold value, quickly evolves into the large oscillations of the upper branch. The inset shows the transition region and indicates that the small charge fronts grow into mature ones. The rate $\lambda$ can also be fitted and now it is positive. The resulting $\lambda$ versus $U$ is plotted in Fig.~\ref{subHopf}(e), showing a linear scaling. This clearly indicates that the bifurcation at $A_l$ is a subcritical Hopf bifurcation. Supercritical Hopf bifurcations in different SL models have been found by Patra \textit{et al.}\cite{PatraS98} and by Hizanidis et al \cite{HIZ05} at low contact conductivity with no shunt. Here we can also see that the time scales have the following relationship: $\tau_b \ll 1/f_{1,2} \ll 1/\lambda$.

It is likely that the bifurcation scenario at $A_u$ in Fig.~\ref{bifurAB}(b) is a saddle-node bifurcation which is probably caused by the collision of the stable limit cycle and the unstable limit cycle that arises from the subcritical Hopf bifurcation at $A_l$. In Fig.~\ref{saddlenode}(c), the power spectrum of the limit cycle ($U=1.4997$~V) and the power spectrum of the transient oscillation at $U=1.49985$~V -  which exceeds the saddle-node bifurcation value $U_{A_u}$ -  are almost identical. This rules out a subcritical torus bifurcation. Then we start the system from a configuration corresponding to the steady oscillation at $U=1.46$~V, but for voltages just above $U_{A_u}$ where there are no limit cycle states, so it eventually reaches the lower branch. Figure~\ref{saddlenode}(a) shows this process at $U=1.49985$~V. After a short time interval of about 1~$\mu$s, the oscillation amplitude $A(t)$ enters a regime of transient oscillations and after a relatively long time $T$, it suddenly exits this region and reaches a steady state. This process looks like a reverse process of Fig.~\ref{subHopf}(d). Figure~\ref{saddlenode}(b) shows the decay of the CALs and CDLs. If we choose the critical value to be 1.499791~V, then the slope in Fig.~\ref{saddlenode}(d) is -0.5. This means that $T \propto \frac{1}{\sqrt{U-U_{A_u}}}$, consistent with a system that undergoes a saddle-node bifurcation of limit cycles.\cite{strogatz}

The bifurcation at point $B$ is at $a=1.00\times10^{-5}$ (Fig.~\ref{bifurB}). For $U \lesssim 2.305$~V, the system oscillates. At first, there is a single CAL in the SL and a dipole is injected from the emitter. The CAL and dipole all move into the SL. The leading CDL moves about twice as fast as the two CALs\cite{AmannS05jsp} and when it catches up with the original CAL, they annihilate. The CAL of the dipole continues to move forward until it reaches the position of the original CAL and stays there for a certain period of time, waiting for another round of dipole injection. Such a bifurcation of a stationary domain state has been reported before by Hizanidis \textit{et al.} \cite{HizanidisB06} for a one-dimensional superlattice model without shunt at higher contact conductivity. The time needed for a dipole to be injected is called the activation time and the time needed to return from the excited state to the fixed point is called the excursion time.\cite{HizanidisB06} As the applied voltage $U$ approaches the boundary, the activation time becomes longer and longer. Taking the critical value $U_{critB}$ of voltage to be 2.30441~V and plotting the frequency of oscillation versus $U-U_{critB}$, we find the frequency obeys the square-root law which is the characteristic scaling law for the saddle-node infinite period bifurcation or SNIPER,\cite{HizanidisB06} which is a global bifurcation of a limit cycle.

Inside the oscillatory region in the $a-U$ parameter space (Fig.~\ref{bifurAB}(a)), we also find regimes of chaos. We still use $a=1.00\times10^{-3}$. As the voltage $U$ decreases inside the oscillatory region, the oscillation shown in Fig.~\ref{osciA} involves a larger part of the SL and the CAL near the emitter becomes less and less prominent until these moving dipoles cover almost all the SL shown Fig.~\ref{chaos}(a),(b) at $U = 1.2$~V and $1.0$~V. Further decrease of the voltage causes the disappearance of the static CAL and the dipoles either annihilate inside the SL or reach and disappear at the collector, shown in Fig.~\ref{chaos}(c) for $U=0.5$~V. Similar chaotic behavior has also been found in SLs without a shunt.\cite{AmannS05jsp} These complicated and apparently chaotic oscillations are found at many points in the oscillatory regime of Fig.~\ref{bifurAB}(a).

In the regime of the stable states between $a\approx 6\times10^{-3}$ and $7\times10^{-6}$ (cf. the right hand region of Fig.~\ref{bifurAB}(a)), the SL usually has a static CAL either inside the SL (for low $a$) or attached to the emitter (for high $a$) and there is a small static CDL to the right of this CAL. This means that the overall field profile is nearly uniform for larger $a$ ($\gtrsim 6\times 10^{-3}$), but static field domains form as $a$ decreases.

\subsection{Dynamical behavior for small contact conductivity}

The bifurcation scenario for lower contact conductivity $\sigma$ is
simpler than for the high $\sigma$ case. Figure \ref{bifurC} shows
the bifurcation diagram for $\sigma = 0.016$~($\Omega $m$^{-1}$).
This value of $\sigma$ corresponds to current self-oscillation in
the SL when there is no shunt.\cite{XuT07} The parameter space is
again divided into an oscillatory regime and a stable stationary
regime. The oscillatory regime starts at about the same value of $a$
as the high $\sigma$ case, i.e., $a\approx 6\times10^{-3}$. However,
this oscillatory region does not have a lower bound. This is because
without the shunt the SL still exhibits oscillations.

The different behaviors at $a=1.00\times10^{-4}$ are shown in Fig.~\ref{bifurC}. As the voltage is deep
inside the oscillatory region, dipoles are periodically injected into the SL and travel through the entire
SL (Fig.~\ref{bifurC}(b)). As voltage increases, the distance that the dipoles travel becomes shorter and
the CAL and CDL annihilate near the emitter (Fig.~\ref{bifurC}(c)). Similar behaviors have been found in
a SL model without shunt.\cite{AmannS05jsp} As the voltage approaches the boundary, the CDL becomes less
and less prominent and the length that the CALs travel becomes even shorter. After the voltage crosses
the boundary, the CALs becomes static. The bifurcation scenario is similar to point A, described in the
previous section, where there is bistability between oscillatory and steady states. The bifurcation
scenarios at other points on the right hand boundary of the oscillatory region appear to be similar to that at point $C$.

\subsection{Dynamical behavior vs. shunt conductivity parameter $b$}

The above discussion focuses only on varying the connectivity parameter $a$ with a shunt of high conductance.
It is also possible to change other parameters of the shunt, such as the conductivity parameter $b$ in the shunt.
A bifurcation diagram can be plotted for $b$ versus $U$ with fixed $a=1.00$ and $\delta_x=200$~nm, and it is
similar to that shown in Fig.~\ref{bifurAB}, with an oscillatory regime between $b \approx 4.5\times10^{-8}$
and $4.5\times10^{-7}$. Another possible control parameter is the width of the shunt. Simulation shows that
only when the width of the shunt is narrower than about $1\times 10^{-4}$~nm, which is unrealistically small,
the SL starts to have oscillation. The oscillatory region for $a$ in Fig.~\ref{bifurAB} and \ref{bifurC} is
almost not affected when $b$ and $\delta_x$ are above certain values so that the current between the shunt
and the SL can always be supported by the shunt. In reality, $\delta_x$ and $b$ should be kept as low as
possible to reduce the power dissipated in the shunt and minimize heat production.

\section{Conclusions}
We have theoretically studied the effect of a shunting side layer
parallel to a semiconductor superlattice, and find that such a
structure can have an almost uniform electric field over the entire
structure even when biased in the negative differential conductivity
(NDC) region. However, even for a shunt with high conductivity and
strong connection to the SL, the field in the SL can be stabilized
only for structures with relatively small lateral extent. As the
lateral size $L_x$ becomes larger, the lateral current in the
quantum well loses the ability to deplete the extra current coming
from the emitter and the field becomes nonuniform. For a
sufficiently thin SL whose lateral dynamics is uniform, the
connection between the shunt and the SL and the conductivity of the
shunt determines the dynamics in the SL. We have also established
the bifurcation diagrams for SLs for different values of the shunt
parameters and identified the presence of both local (Hopf) and
global (SNIPER) bifurcations.

Although the microscopic nature of electronic transport in
weakly-coupled SLs is different than for strongly-coupled SLs, the
NDC property is known to produce similar dynamics in both types of
structures when they are not shunted. Thus, it seems plausible that
for suitable shunt connectivity and SL lateral width that a
stronly-coupled SL might also be stabilized with a shunting side
layer. This could enable the realization of a SL-based THz
oscillator.

\begin{acknowledgements}
This work was supported in part by NSF grant DMR-0804232, by IRCSET, and by DFG in the framework of Sfb 555.
\end{acknowledgements}

\appendix*
\section{Numerical method}
In order to implement the implicit method, the dynamical variables, i.e. the electron densities $n_m(x)$, should be computed from the system Eqs.~(\ref{SLdynamics}) and (\ref{shuntdynamics}). However, $n_m(x)$ are deeply buried in these equations, where the currents depend on the field that relates to $n_m(x)$ by solving Poisson equation Eq.~(\ref{Poisson}). So instead of solving for $n_m(x)$ directly, we use the semi-implicit Euler method and numerically calculate the Jacobian matrix that is needed for this method. The procedure is as follows: after discretization of the space, the quantities of potential and charge density are placed on the grid. The fields and currents (also the charge density that is needed to calculate the currents) are placed on a staggered grid. Knowing the charge density distribution, the potential is determined by the Poisson equation using a method described in Ref.~\onlinecite{AmannS05}. After that, the currents to each grid point are calculated from the electric fields which are immediately obtained from the potential (cf. Eqs.~(\ref{SLfield}) and (\ref{shuntfield})). Then the charge densities are iterated one step forward in time as
\begin{equation}
e\mathbf{n}'=e\mathbf{n}+dt\mathbf{J}(\mathbf{n}'),
\end{equation}
where $\mathbf{n}=(n_{11}, n_{12}, ..., n_{21}, n_{22}, ...)^T$ is the vector whose components are the charge densities on each grid point. The first subscript denotes the SL period number and the second one is the grid point index in the $x$ direction. The vector current $\mathbf{J}$ is the total current flow into or out of each grid point. $\mathbf{n}'$ is the new charge density configuration after time step $dt$. Since we are using the implicit method, $\mathbf{J}$ must depend on the future charge density configuration instead of the old one. We linearize the equations:
\begin{equation}
\mathbf{n}'=\mathbf{n}+dt\left[\mathbf{J}(\mathbf{n}) + \frac{\partial \mathbf{J}}{\partial \mathbf{n}}\bigg|_{\mathbf{n}}\cdot(\mathbf{n}'-\mathbf{n})\right],
\end{equation}
where $\partial J/\partial n$ is the Jacobian matrix. Rearranging this equation yields:
\begin{equation}
\mathbf{n}'=\mathbf{n}+ dt\left[ \mathbf{1}-dt\frac{\partial \mathbf{J}}{\partial \mathbf{n}}\right]^{-1}\cdot \mathbf{J}(\mathbf{n})
\label{A3}
\end{equation}
We mentioned that the currents do not depend on the charge densities explicitly. So to calculate the Jacobian matrix, we first calculate $\mathbf{J}(\mathbf{n})$, then slightly change the charge density at one grid point to $n_{ij}+\delta n_{ij}$ and calculate the currents $\mathbf{J}'$ based on this charge configuration. Then one row of the Jacobian matrix is immediately obtained by $(\mathbf{J}'-\mathbf{J}(\mathbf{n}))/\delta n_{ij}$.

To solve Eq.~(\ref{A3}), we do not invert the matrix. Instead, we write it as:
\begin{equation}
dt \mathbf{J}(\mathbf{n}) =  \left[ \mathbf{1}-dt\frac{\partial \mathbf{J}}{\partial \mathbf{n}}\right]\cdot(\mathbf{n}'-\mathbf{n}).
\end{equation}
Then we solve this set of linear equations by Gauss elimination.


\begin{thebibliography}{10}

\bibitem{EsakiT70}
L. Esaki and R. Tsu, IBM J. Res. Develop {\bf 14},  61  (1970).

\bibitem{Bloch28}
F. Bloch, Z. Phys. {\bf 52},  555  (1928).

\bibitem{Zener34}
C. Zener, Proc. R. Soc. London, Ser A {\bf 145},  523  (1934).

\bibitem{KtitorovS72}
S.~A. Ktitorov, G.~S. Simin, and V.~Y. Sindalovskii, Sov. Phys. Solid State {\bf 13},  1872  (1972).

\bibitem{FeldmannL92}
J. Feldmann, K. Leo, J. Shah, D.~A.~B. Miller, and J.~E. Cunningham, Phys. Rev. B {\bf 46},  7252  (1992).

\bibitem{MendezA88}
E.~E. Mendez, F. Agull{\'{o}}-Rueda, and J.~M. Hong, Phys. Rev. Lett. {\bf 60}, 2426  (1988).

\bibitem{YajimaT79}
T. Yajima and Y. Taira, J. Phy. Soc. Jpn. {\bf 47},  1620  (1979).

\bibitem{LeoS91}
K. Leo, J. Shah, E.~O. G{\"{o}bel}, T.~C. Damen, S. Schmitt-Rink, W.
  Sch{\"{a}}fer, and K. K{\"{o}}hler, Phys. Rev. Lett. {\bf 66},  201  (1991).

\bibitem{Gunn63}
J.~B. Gunn, Solid Stat. Commun. {\bf 1},  88  (1963).

\bibitem{SchomburgH97}
E. Schomburg, K. Hofbeck, J. Grenzer, T. Blomeier, A.~A. Ignatov, K.~F. Renk,
  D.~G. Pavel'ev, Y. Koschurinov, V. Ustinov, A. Zhukov, S. Ivanov, and P.~S.
  Kop'ev, Appl. Phys. Lett. {\bf 71},  401  (1997).

\bibitem{WilliamsK05}
B.~S. Williams, S. Kumar, Q. Hu, and J.~L. Reno, Opt. Express {\bf 13},  3331 (2005).

\bibitem{WillenbergD03}
H. Willenberg, G.~H. Dohler, and J. Faist, Phys. Rev. B {\bf 67},  085315 (2003).

\bibitem{THz}
Phil. Trans. Roy. Soc. Lond. A., {\bf 362}, (2004) special issue ``THz-gap''.

\bibitem{HyartA08}
T. Hyart, K.~N. Alekseev, and E.~V. Thuneberg, Phys. Rev. B {\bf 77},  165330 (2008).

\bibitem{Kroemer}
H. Kroemer, arXiv:cond-mat/0009311  .

\bibitem{SavvidisL04}
P.~G. Savvidis, B. Kolasa, G. Lee, and S.~J. Allen, Phys. Rev. Lett. {\bf 92}, 196802  (2004).

\bibitem{BaoW06}
M. Bao and K.~L. Wang, IEEE Trans. Electron Devices {\bf 53},  2564  (2006).

\bibitem{WallmarkV63}
J.~T. Wallmark, L. Varettoni, and H. Ur, IEEE Trans. Electron Devices {\bf 10}, 215  (1963).

\bibitem{DanielG05}
E. Daniel, B. Gilbert, J. Scott, and S. Allen, IEEE Trans. Electron Devices {\bf 50},  2434  (2003).

\bibitem{BonillaG05}
L.~L. Bonilla and H.~T. Grahn, Rep. Prog. Phys {\bf 68},  577  (2005).

\bibitem{FeilT05}
T. Feil, H.-P. Tranitz, M. Reinwald, and W. Wegscheider, Appl. Phys. Lett. {\bf 87},  212112  (2005).

\bibitem{Wacker02}
A. Wacker, Phys. Rep. {\bf 357},  1  (2002).

\bibitem{SCH01}
E. Sch{\"o}ll, \textit{Nonlinear spatio-temporal dynamics and chaos in semiconductors} (Cambridge University Press, Cambridge, 2001), {Nonlinear Science Series}, Vol. 10.

\bibitem{XuT07}
H. Xu and S. Teitsworth, Phys. Rev. B {\bf 76},  235302  (2007).

\bibitem{AmannS05}
A. Amann and E. Sch{\"{o}}ll, Phys. Rev. B {\bf 72},  165319  (2005).

\bibitem{CHE00}
V. Cheianov, P. Rodin, and E. Sch{\"o}ll, Phys.~Rev.~B {\bf 62},  9966  (2000).

\bibitem{Xuu}
H. Xu, unpublished.

\bibitem{AMA02}
A. Amann, A. Wacker, and E. Sch{\"o}ll, Physica~B {\bf 314},  404  (2002).

\bibitem{AMA02a}
A. Amann, J. Schlesner, A. Wacker, and E. Sch{\"o}ll, Phys.~Rev.~B {\bf 65}, 193313  (2002).

\bibitem{PatraS98}
M. Patra, G. Schwarz, and E. Sch{\"o}ll, Phys. Rev. B {\bf 57},  1824  (1998).

\bibitem{HIZ05}
J. Hizanidis, A.~G. Balanov, A. Amann, and E. Sch{\"o}ll, Int.~J.~Bifur.~Chaos {\bf 16},  1701  (2006).

\bibitem{strogatz}
S.~H. Strogatz, \textit{Nonlinear Dynamics and Chaos: With Applications to Physics, Biology, Chemistry and Engineering} (Westview Press, New York, 2001).

\bibitem{AmannS05jsp}
A. Amann and E. Sch{\"{o}}ll, J. Stat. Phys {\bf 119},  1069  (2005).

\bibitem{HizanidisB06}
J. Hizanidis, A. Balanov, A. Amann, and E. Sch{\"{o}}ll, Phys. Rev. Lett. {\bf  96},  244104  (2006).

\end{thebibliography}
\end{document}